\documentclass[prl,nofootinbib,twocolumn,superscriptaddress]{revtex4}

\def\mysection#1{{\bf #1.} }

\usepackage{amssymb}
\usepackage{amsmath}
\usepackage[dvips]{graphicx}
\usepackage{longtable}
\usepackage{verbatim}
\usepackage{amsfonts}

\newcommand{\C}{\mathbb{C}}
\newcommand{\Z}{\mathbb{Z}}

\newcommand{\ten}{\mathbf{10}}
\newcommand{\bfive}{\mathbf{\bar 5}}

\begin{document}
{\hspace*{13cm}\vbox{\hbox{WIS/04/07-MARCH-DPP}}}

\title{Dynamical Supersymmetry Breaking from Simple Quivers}

\author{Yaron E. Antebi}\email{yaron.antebi@weizmann.ac.il}
\affiliation{Department of Particle Physics,
  Weizmann Institute of Science, Rehovot 76100,
  Israel}

\author{Tomer Volansky}\email{tomer.volansky@weizmann.ac.il}
\affiliation{Department of Particle Physics,
  Weizmann Institute of Science, Rehovot 76100, Israel}

\begin{abstract}
  We construct a simple local model of dynamical supersymmetry
  breaking.  The model is a one generation $SU(5)$ that arises from a
  IIB $\Z_N$ orientifold.  It does not admit a runaway direction and
  is argued to stabilize the blowup mode related to the corresponding
  $U(1)$ factor.  The theory demonstrates the existence of a new class
  of ``blowup'' fractional branes.  
\end{abstract}

\maketitle

\section{Introduction}
\label{sec:introduction-1}

Dynamical supersymmetry breaking (DSB) \cite{Witten:1981nf} is an
intriguing solution to the hierarchy problem.  Examples of such models
were first presented more than twenty years ago \cite{Affleck:1984xz},
and the idea has been extensively studied both from the theoretical
and phenomenological points of view (for a review see e.g.
\cite{Shadmi:1999jy}).

For string theory to make contact with reality, some mechanism to
break supersymmetry must be employed.  In recent years, following the
understanding of flux compactifications and moduli stabilization
\cite{Giddings:2001yu,Kachru:2003aw}, the problem of breaking
supersymmetry has attracted a vast amount of attention.  Many models
have been presented, employing various stringy mechanisms, however only
very few break supersymmetry dynamically.  The reason for the lack of 
DSB models in string theory is twofold.  On the one hand,
models with completely stable DSB vacua are non-generic at the field
theory level. On the other hand, compactifying such models and
taking care of the stabilization of all moduli and in particular
K\"ahler moduli, is very laborious \cite{Diaconescu:2005pc}.  

Constructions of local models were attempted on D-branes
\cite{Lykken:1998ec}.  In
\cite{Franco:2005zu,Berenstein:2005xa,Bertolini:2005di}, a
classification of the gauge dynamics on fractional branes was
introduced, where it was argued that the corresponding quiver theories
typically break supersymmetry dynamically.  However, as was stressed
in \cite{Intriligator:2005aw}, these brane configurations generically
posses a runaway direction which corresponds to a blowup of the
singular geometry.  This problem can be ameliorated in compact models
by stabilizing the runaway directions through some non-perturbative
effects \cite{Florea:2006si,Diaconescu:2007ah}.

Recently it was suggested that meta-stable vacua that exhibit DSB may
be more generic \cite{Intriligator:2006dd}.  While indeed true at the
field theory level, such constructions in string theory still lack a
good explanation for the origin of small mass terms which appear in
most theories.  There have been several attempts to realize such
models in string theory, however most do not address the above
issue \cite{Ooguri:2006bg,Franco:2006es,Argurio:2006ny} and cannot be
compactified in a direct manner.

It is therefore worthwhile to construct new local models of DSB which
are simple enough to allow for a straightforward embedding in a
compact background.  In this note we report on progress in this
direction.  Here we concentrate on a simple local construction, while
the details of the compact model will be given in \cite{avFull}. The
local construction is a type IIB $\Z_N$ orientifold.  Specifically, we
construct an $SU(5)$ gauge theory with one generation of $\ten+\bfive$
\cite{Affleck:1983mk}.  After imposing the orientifold projection,
only one anomalous $U(1)$ is present.  We argue that the corresponding
closed string K\"ahler blowup mode that shows up as a Fayet-Iliopoulos
(FI) term is stabilized close to the origin.  As opposed to the
generic quiver, this (bi-directional) quiver does not suffer from a
runaway behaviour and demonstrates the existence of a new class of
fractional branes which we call {\it blowup fractional branes}.

While this work was being completed, we became aware of
\cite{Wijnholt:2007vn} which partially overlaps with the local
construction of our model.

\section{Local Models}
\label{sec:local-models}

As a first step towards writing a complete compact solution, one must
specify a local quiver model which exhibits DSB.  Here we concentrate
on the non-calculable $SU(5)$ gauge theory with one generation of
$\ten$ and $\bfive$.  This model is known to break supersymmetry
dynamically \cite{Affleck:1983mk}.  The
construction is based on fractional branes located at fixed points of
$\C^3/\Z_N$ orientifolds.  Quiver models that arise from placing
D-branes at such singularities have been extensively studied.  The
reader is referred to \cite{Kiritsis:2003mc,Aldazabal:2000sa} and
references therein for more details.

\mysection{Formalism} We begin by setting up our notations, closely
following \cite{Aldazabal:2000sa}. Consider a $\C^3/\Z_N$ singularity.
The $\Z_N$ generator $\theta$ acts on the three complex coordinates as
$\theta: (z^1,z^2,z^3) \rightarrow (\omega^{b_1}z^1,
\omega^{b_2}z^2,\omega^{b_3}z^3)$ where $\omega=e^{2\pi i/N}$ is the $N$th root
of unity.  To preserve ${\cal N}=1$ supersymmetry, the $\Z_N$ must be
a subgroup of $SU(3)$ which translates into taking $b_1+b_2+b_3=0$
(mod $N$).  The action on the Chan-Paton indices is
\begin{eqnarray}
  \label{eq:73}
   A^\mu &\rightarrow& \gamma(\theta)A^\mu\gamma(\theta)^{-1},
  \\
  Z^i &\rightarrow& \omega^{b_i}\gamma(\theta)Z^i\gamma(\theta)^{-1},
\end{eqnarray}
where $\gamma(\theta)$ is a representation of $\Z_N$.  Since $\Z_N$ is
abelian, all its irreducible representations are one dimensional, and
without loss of generality we may take $\gamma(\theta)$ to be
\begin{equation}
  \label{eq:74}
  \gamma(\theta) = \text{diag}({\bf 1}_{n_0}, \omega {\bf 1}_{n_1},
  ... ,\omega^{N-1}{\bf 1}_{n_{N-1}})
\end{equation}
where $\sum_a n_a = n$ is the number of fractional branes.  The
invariant spectrum at the singularity is described by a $U(n_0)\times
U(n_1)\times...\times U(n_{N-1})$ theory with matter multiplets
transforming as $({\bf n_a},{\bf \bar n_{a+b_i}})$ for $i = 1,2,3$ and
$a+b_i$ is taken mod $N$.  Such a field theory can be efficiently
described by a quiver diagram, where each node denotes a $U(n)$ factor
and the bi-fundamental chiral fields are represented by directed lines
connecting two such nodes. A line originating and ending on the same
node describes a field in the adjoint representation of the
corresponding $U(n)$ factor.

Next we would like to consider the spectrum of D-branes located on top
of orientifold planes.  As usual, to preserve the same supersymmetry
as $D3$ branes, only $O3$- or $O7$-planes may be included, located on
the fixed locus of the orientifold action, $\Omega R (-1)^{F_L}$
(where $R$ is the $Z_2$ geometric involution and $(-1)^{F_L}$ is the
left-handed world-sheet fermion number).  In terms of the open string
modes, the effect of the orientifold action is to identify each
$U(n_a)$ gauge group with $U(n_{-a})$ while identifying the representation
$({\bf n_a},{\bf \bar n_{a+b_i}})$ with $({\bf n_{-a-b_i}},{\bf \bar
  n_{-a}})$.  In particular, the $U(n_0)$ and $U(n_{N/2})$ gauge
factors (if exist) are projected onto themselves, resulting in an
$Sp$($SO$) gauge group, depending on the exact orientifold action.
Similarly chiral fields transforming in the $({\bf n_a},{\bf \bar
  n_{-a}})$ are projected into the (anti)symmetric representation of
$SU(n_a)$.

Finaly, the quiver diagrams must be extended to accommodate these
unoriented theories \cite{Antebi:2005hr}.  Since each end of the string can
independently be in either the fundamental or the antifundamental, it
must be represented as a bi-directed line with an arrow at each of the
two ends, indicating the representation of the string under
each of the two gauge group factors. In such a bidirected quiver
(biquiver for short), a symmetric or an antisymmetric field is
represented by a line with both ends coming out of the same set of
branes.

\begin{figure}[tb]
  \centering
  \includegraphics[scale=0.6]{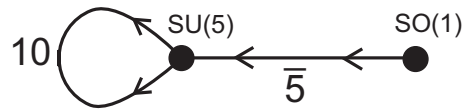}
  \caption{A one-generation SU(5) Biquiver.}
  \label{fig:Quiver_SU5}
\end{figure}

\mysection{DSB Quivers} It is now a simple matter to construct the
desired $SU(5)$ model.  As an example, consider a $\Z_6$ orientifold
with the orbifold action $(b_1,b_2,b_3) = (1,2,-3)$ and orientifold $R
=(-1,-1,-1)$.  Furthermore, we take the action on the Chan-Panton
indices to be,
\begin{eqnarray}
  \label{eq:75}
  {\rm diag}({\bf 1}_1,\omega^2 {\bf 1}_5,\omega^{4} {\bf 1}_5)
  ,\qquad \omega=e^{i\pi/3},
\end{eqnarray}
so altogether we have eleven fractional branes.  There is a single
orbifold fixed point and an $O3$-plane at the origin.  The biquiver is
shown in Fig. \ref{fig:Quiver_SU5}.
As required, the theory is $SO(1)\times U(5)$ with one generation of
$\ten + \bfive$.  The $U(1)$ corresponding to the $SU(5)$ is anomalous
and becomes massive through the generalized Green-Schwarz
mechanism \cite{Green:1984sg,Dine:1987xk}. Hence it remains as a
global symmetry and has no effect on the low-energy dynamics.

It is also possible to construct this biquiver on other singularites,
as we (non-exhaustively) list in table \ref{tab:su5models}.

\begin{table}[tb]
  \centering
  \begin{tabular}{|c|c|c|}
    \hline
    N & $b_i$ & $\gamma(\theta)$\\
    \hline
    \hline
    $\Z_6$ & (1,2,-3) & ${\rm diag}({\bf 1}_1,\omega^2 {\bf 1}_5,\omega^{4} {\bf 1}_5)$\\
    \hline
    $\Z_6$ & (1,1,-2) & ${\rm diag}({\bf 1}_1,\omega^2 {\bf 1}_5,\omega^{4} {\bf 1}_5)$\\
    \hline
    $\Z_9$ & (2,4,-6) & ${\rm diag}({\bf 1}_1,\omega^3 {\bf 1}_5,\omega^{6} {\bf 1}_5,)$\\
    \hline
    $\Z_{12}$ & (1,4,-5) & ${\rm diag}({\bf 1}_1,\omega^4 {\bf 1}_5,\omega^{8} {\bf 1}_5)$\\
    \hline
  \end{tabular}
  \caption{The four different orientifold models giving rise to the non-calculable $SU(5)$ DSB theory.}
  \label{tab:su5models}
\end{table}

\mysection{Blowup Fractional Branes} Such DSB biquivers exhibit a new
class of fractional brane models.  In \cite{Franco:2005zu} fractional
branes were classified as follows:
\begin{enumerate}
\item {\it {\cal N}=2 fractional branes:}  These exhibit flat
  directions along which the dynamics are those of ${\cal N}=2$.  They
  typically arise at non-isolated $\C^2/\Z_N$ singularities.
\item {\it Deformation fractional branes:} The 
  theory exhibits confining dynamics which translates into (partial) complex
  structure deformation of the geometry.  
\item {\it Runaway (DSB) fractional branes:} This is the generic
  case.  In general, the gauge factors have different ranks and the
  dynamics lead to a runaway behaviour through a non-perturbative
  superpotential. 
\end{enumerate}
 
It is clear that the DSB biquiver described above does not fit into
any of the above classes but instead demonstrates the existence of a
new class of fractional branes:\footnote{We thank Angel Uranga for
  drawing our attention to this point.}
\begin{enumerate}
\item[4.] {\it Blowup fractional branes:} These are fractional branes
  which do not have flat or runaway directions and are associated with
  the stabilization of K\"ahler moduli, corresponding to the possible
  (partial) blowup of the singularity.  In accordance with the
  classification above we expect blowup fractional branes to be
  related to unoriented singularities.
\end{enumerate}
In our example the singularity indeed blows up, as we now explain.
Ignoring for the moment the non-perturbative dynamics, the $SU(5)$
model has a classical supersymmetric minimum at the origin of field
space.  Turning on a FI term $\xi^2$ for the corresponding $U(1)$,
breaks supersymmetry due to the incompatibility between the $SU(5)$
and $U(1)$ D-terms.  At large $\xi^2$ where the classical theory is
reliable, a potential
\begin{eqnarray}
  \label{eq:76}
  V \sim |\xi|^4
\end{eqnarray}
is generated, driving the dynamical FI field to zero.  Taking the
non-perturbative effects into account, one cannot determine the exact
location of the minimum, and on dimensional grounds we expect $\xi^2$
to stabilize near the origin at $\xi^2 \sim \Lambda_5^2$.  Such
stabilization corresponds to blowing up a 2-cycle in the geometry.

This is in contrast to the case of the runaway class, for which the
D-term of a massive anomalous $U(1)$ is necessary in order to
stabilize a classical flat direction that becomes unstable quantum
mechanically.
However, as was already noted in \cite{Franco:2005zu} and stressed in
\cite{Intriligator:2005aw}, such D-term equations should not be
imposed, as the massive $U(1)$ is not exhibited at low energy.
Imposing the massive D-terms comes at the expense of introducing a new
runaway direction of a blowup mode which appears as a FI term.  For
the model at hand, the field theory does not have a runaway direction
and this, in turn, translates into having a stabilized K\"ahler
modulus.

For the specific $SU(5)$ local model, one encounters at the field
theory level a single FI blowup mode.  In order to embed this quiver
in a compact model (away from the decoupling limit), one must worry
about other K\"ahler moduli which must be stabilized without changing
the theory at the singularity.  There are two  mechanisms: First,
for the specific $\Z_6$ orbifold, the local geometry consists of four
exceptional divisors (arising from one fixed-point and two
fixed-curves) out of which only one is compact.  Thus, out of the four
twisted K\"ahler moduli, one is stabilized as seen through the gauge
dynamics, while the others may  be stabilized away from the
orbifold fixed-point without affecting the quiver.  Second, it is not
at all clear which (if any) of the K\"ahler moduli remain after the
orientifold projection.  It is possibly misleading to understand the
geometry by first resolving the singularity and then orientifolding.
Still, the analysis of \cite{Lust:2006zh} suggests that at least some
of these moduli might be projected out.  More details of the K\"ahler
stabilization will appear in \cite{avFull}.

Finally, let us remark that at this stage it is still not clear
how generic the blowup class is or whether examples exist where the
K\"ahler moduli are stabilized exactly at the origin, corresponding
to the orbifold limit.  Furthermore, it would be very interesting to
understand whether such quivers exhibit a large-$N$ limit with DSB and
K\"ahler stabilization at the bottom of a duality cascade.  We postpone
the investigation of this question to future work.

\section{Summary}
\label{sec:summary}

In this letter, a novel realization of the one generation $SU(5)$ DSB
model in string theory was introduced. The model arises in a simple
type IIB $\Z_N$ orientifold with fractional branes at the singular
locus.  The corresponding biquiver model is easily extracted from the
geometry. At the field theory level the model has no flat
directions, which translates into a stabilization of the K\"ahler
modulus.  The latter appears as a dynamical FI term related to the
anomalous $U(1)$.  The dynamics are therefore in a new class of
fractional branes, which (partially) blow up the geometry.

Such models are very simple and are generated from singularities that
appear generically on the moduli space of Calabi-Yau manifolds.
Therefore it should be easy to construct a compactify version of our
construction embedding it in a CY 3-fold. It will be interesting to
further study such constructions as they will allow for
complex-structure moduli stabilization by turning on fluxes. Such a
setup is a step forward in constructing realistic models of particle
physics and may allow one to address issues of DSB in the landscape.

\section{Acknowledgments} 
We thank O. Aharony, G. Engelhard, Y. Hochberg, S. Kachru, Z.
Komargodski, Y. Shadmi, J. Simon and A. Uranga for useful discussion.
We want to specially thank M. Berkooz, M. Dine, and B. Florea
for many valueable and important discussions.
This work was supported in part by the Israel-U.S.  Binational Science
Foundation, by the Israel Science Foundation, by the European network
HPRN-CT-2000-00122, by a grant from the G.I.F., the German-Israeli
Foundation for Scientific Research and Development, and by a grant of
DIP (H.52).
TV would like to thank the KITP for its hospitality during the course
of this project.  YEA would like to thank Stanford University and
SLAC for their hospitality during the course of this project.

\end{document}